\documentclass[12pt]{article}
\usepackage{amstex,amscd}
\newcommand{\be}{\begin{eqnarray}}
\newcommand{\ee}{\end{eqnarray}}
\newcommand{\beq}{\begin{equation}}
\newcommand{\eeq}{\end{equation}}
\newcommand{\abs}[1]{\left| \! #1 \! \right|}

\newcommand{\C}{C\hspace{-0.7em}I\:}
\newcommand{\R}{R\hspace{-1em}I\ \ }
\newcommand{\I}{I\hspace{-0.4em}I\:}
\newcommand{\1}{1\hspace{-0.3em}\rm I \:}
\newcommand{\M}{M\hspace{-1.2em}I\ \: }

\newcommand{\Spa}{S\hspace{-0.6em}I\: }
\newcommand{\T}{T\hspace{-0.8em}I\ }

\newcommand{\sR}{R\hspace{-0.7em}I\ }
\newcommand{\lra}{\longrightarrow}
\newcommand{\Li}{L\hspace{-0.8em}I\ }

\begin{document}

\thispagestyle{empty}
\hfill MPI-PhT/97-54 

\vspace{3ex} 

\begin{center}

{\Huge General Aspects of Symmetry Breaking  \\}
{\large
\vspace{4ex} 
H. Martin Haft \\
{\small mah$@$mppmu.mpg.de}\\
\vspace{5ex}
{\it Max-Planck-Institut f\"{u}r Physik \\
 (Werner-Heisenberg-Institut) \\
 80805 Munich \\
 Germany }
}

\end{center}

\vspace{5ex}
\centerline{\bf Abstract}

On basis of an algebraic analysis of symmetry breaking in general and the
Higgs mechanism in the standard model of elementary particles we generalize 
the concept of symmetry breaking to systems with non-compact groups but 
not necessarily caused by a potential. Thereto we give some simple, but
unfamiliar examples of symmetry breaking with and without potentials. The
analysis of the concept of mass in space-time and in the Higgs mechanism 
will lead to a model unifying both structures in terms of symmetry breaking 
of $GL(2,\C)$.



\newpage
\pagenumbering{arabic}

\subsection*{Introduction}

In his famous work of 1916 \cite{Ein} A.~Einstein motivated the necessity of an
extention of special relativity to general relativity with the help of a
gedankenexperiment based on Mach's principle. It states that it is not 
possible to experience the motion of a single isolated object in space-time 
without reverence to other objects. Einstein concluded that the absolute 
Galileian or Newtonian
space-time, and equally the concept of space-time in special relativity, is
just an ``imaginary cause'' (fiktive Ursache \cite{Ein}). 
His postulate of general covariance states that there
is no given absolute basis in space-time, i.e.~no given reverence frame, 
and therewith no given form of the metric, but
the local bases and therewith the local metrical structure is due to the 
mass distribution in space-time. 

Pushing Mach's principle to the extreme we have to state that we can't
experience space-time without any objects to test space-time 
(also fields, like the
electro-magnetic field, have to be tested with the help of objects), and we
can't define length or volume, mass or energy and (angular) momentum without
reverence objects. On the other hand objects themselves are always regarded 
as objects in space-time. Hence this serves for a paradoxy. 

Where does our concept of a particle (object) and its mass come from? 
Firstly, the definition of a particle in special relativity is given by 
Wigner \cite{Wig PG}, particles being positive unitary irreducible 
representations of the Poincar\'e group. Therewith mass is the Casimir 
invariant of the translations within the Poincar\'e group and therewith 
the abstract concept of inertial mass. Modern quantum field theory, 
especially the standard model of elementary particles \cite{Wei}, has given 
another approach to the concept of mass, connected with the Higgs mechanism 
\cite{Hig}. The Higgs mechanism creates in a gauge invariant way at least 
formally terms within the Lagrange densities, which are already known to be 
mass terms for particles, leading to already known equation of motions for 
massive particles. There is no conceptual theory why these `masses', 
generated out of an internal symmetry structure, are equal to the external 
concept of mass according to the Poincar\'e group. 
Even more there is no conceptual
connection to the role of mass in general relativity. 

The mass terms in the standard model are generated by the concept of
spontaneous symmetry breaking. With the above problems in mind we analyze in
this work the general structures of symmetry breaking. This is done with the
aim to get rid of the existence of an a priori given potential, 
which already presupposes the concept of energy. Because of the
non-compact structure of the external space-time symmetry this analysis is 
also
performed to generalize the concept of symmetry breaking of compact groups 
to
non-compact groups. Based on this analysis, specified by some simple
examples, we propose a quite abstract model which has the potential to 
resolve the paradoxy and explain the duality of space-time and objects in 
terms of symmetry breaking.

\subsection*{Algebraic Aspects of Symmetry Breaking}
Spontaneous symmetry breaking is an important feature of the standard model 
of elementary particles as well as for grand unified theories. Besides other
features it is essential to
define the observed symmetry patterns and to generate in a gauge invariant 
way mass terms for gauge bosons and fermions. To name the basic aspects of 
symmetry breaking the Higgs sector of the standard 
model will be analyzed shortly: Omitting the kinetic
terms, the Higgs sector of the standard model is given by an 
$U(2)$-symmetric
potential on a complex $2$-dimensional vector space $V$ of the Higgs field
$\Phi (x)$: 
\[ 
	\Phi (x) \in V \cong \C^2, \ x \in \M 
\]
\beq
	V_{\mbox{Higgs}}(\Phi) \ \ = \ \mu^2 \Phi^* \Phi + 
		\lambda (\Phi^* \Phi)^2 , \ \ \ \lambda > 0 .\label{HPot}
\end{equation}
For $\mu^2 > 0$ no symmetry breaking occurs, i.e. the ground state (state of
lowest potential energy) is unique and invariant under the action of the 
whole symmetry group $U(2)$. For $\mu^2 < 0$ the possible ground states are 
only invariant under the action of a $U(1)$ subgroup of $U(2)$. The 
manifold of all possible ground states is given by the $3$-parametric 
(Goldstone) coset space
$U(2) / U(1)$. This situation is often exemplified in $\R^2$. Take the 
$\R^2$-dimensional analoque potential with $O(2)$-symmetry: 
\[ 
	y \in  \R^2 
\]
\beq
	V(y) \ \ = \ a y^2 + b (y^2)^2 , \ \ \ b > 0 .\label{MHPot}
\end{equation}
For $a > 0$ the ground state is unique, given by the origin, with complete 
symmetry $O(2)$. For $a < 0$ (Mexican hat potential) 
there is a $\$^1$-sphere (circle) of ground states
defined by the vectors with length $\abs{y} = \sqrt{\frac{-a}{2b}} =: R$. 
Contrary to the Higgs
potential the invariance groups of these vectors are discrete. The
manifold of ground states is given with $O(2) / \I_2  \cong SO(2) 
\cong \$^1$ (with $\I_2 \cong \{ \1_2, {\tiny \left( \begin{array}{cc} 1 & 0 \\ 
0 & -1  \end{array} \right)} \} $ in a certain basis).

To distinguish between $a > 0$ and $a < 0$ is like to distinguish between the
different strata of the action of $O(2)$ on $\R^2$. A stratum is the collection
of all isomorphic orbits of the action of a group. In general: The action of a
group $G$ on a vector $v$ of a representation space $V$ of this group defines 
the orbit $[ v ]_G$ of this vector with respect to the representation $\rho$ 
of the group: 
\begin{equation} 
	\begin{array}{rl}
		\mbox{representation}: & \rho \ : \ G \lra \mbox{end}(V);  \\ 
		\mbox{endomorphism}: & \rho(g) \ : \ V \lra V , \ \ g \in G ; 
			\\ 
		\mbox{$G$-action}: & \rho(g)(v) \ =: \ g \bullet v \ \in \ V ; 
			\\
		\mbox{orbit}: & [ v ]_G \ := \ \rho(G)(v) \ = \ G \bullet v.
	\end{array} \nonumber
\end{equation}
$V$ is thus decomposed non-linearily into orbits of the action of $G$. 
There may be a non-trivial subgroup $H_v \subset G$ which leaves the vector 
$v$ invariant, i.e. 
\[
	H_v \bullet v \ = \ v .
\]
This subgroup is called fixgroup or isotropy group of $v$ or, 
due to Wigner \cite{Wig PG}, little
group. Regarded as manifold the orbit $[v]_G$ is isomorphic to the coset
(or homogeneous) space $G/H_v$,
\[
	[v]_G \ \cong \ G/H_v .
\]
If $H_v$ is a normal subgroup this manifold has in addition a group structure. 
If $v$ and $w$ are two different elements of the same orbit there is at least 
one $g \in G$
with $w = g \bullet v$. Therewith $v$ and $w$ have isomorphic but in general 
not the same fixgroups:
\[
	H_w \ = \ g H_v g^{-1} \ .
\]
I.e.~the fixgroups of elements of one orbit are conjugated. 
Thus the concept of a little group
is a `local' structure on the manifold $G/H$: its representation depends on the
element of the manifold. The collection of all orbits with isomorphic 
fixgroups, i.e.~the same fixgroup up to conjugation, is called a stratum. 
Thus $V$ is 
decomposed non-linearily into strata with smaller decomposition given by the
orbits. 

With this notation the two examples for spontaneous symmetry breaking given
above can be characterized as follows: 
A $G$-symmetric potential defines via its ground
states one orbit within a stratum. In case of the little group of this orbit 
being isomorphic to the whole group $H \cong G$ the orbit is trivial, 
$[ v ]_G \cong G/G \cong \{ 1 \} $, i.e.~the ground state is unique. 
Otherwise we have a non-trivial orbit and therewith a non-trivial manifold of 
ground states. In the example of $O(2)$ acting on $\R^2$ the non-trivial 
stratum is
the collection of all concentric circles around the origin. These orbits within
the stratum are connected with a dilatation operation, i.e.~by the action of
the group $D(1) = e^{\sR} \cong \R^+$. 
For the non-trivial stratum the selection of an
orbit within the stratum introduces a length scale, $\abs{y} = R$ 
(which in the Higgs model becomes an energy or mass scale), 
and therefore breaks the $D(1)$
invariance of the stratum. I.e.~hidden in the potential there is an explicit
breaking of the $D(1)$ structure of the stratum by an explicit introduction of 
a scale, e.g.~$\sqrt{\frac{-a}{2b}}$ or $\sqrt{\frac{- \mu^2}{2 \lambda}}$. 
Spontaneous symmetry breaking now distinguishes one
vector within the orbit. For 
all of these vectors the action of the group $G$ on this vector 
generates (per definition) the
orbit. The action of the group together with the action of $D(1)$
generates the whole stratum out of one vector. 

In this context spontaneous symmetry breaking has two features: First
fixing a scale within the stratum and secondly distinguishing one vector which
has the possibility to give one basis vector for the representation space.
Together this will provide one normal basis vector of the representation space
of $G$. Notice, within a vector space there is normally given no natural,
i.e.~preferred, basis. However, mostly we are used to work in a basis. On the
other side general relativity teaches to understand nature basis-free as long as
nature itself doesn't distinguish a basis.  
It is the distinction of one preferred (basis) vector which causes
symmetry breaking. To ask for symmetry breaking is first of all to ask for the
distinction of a non-trivial vector within a representation space of a symmetry
group. Therefore the describtion given above focuses on a algebraic and
therewith basis-free describtion of symmetry breaking. We will stress this point
of view in the examples given below. 

Of course the algebraic description of symmetry breaking in terms of orbits or
cosets and little groups is well known, see e.g.~\cite{OR}. However, these
structures are only used in connection with compact symmetry groups $G$ due to
the assumption of the existence of potentials like (\ref{HPot}) or
(\ref{MHPot}). Here we emphasized certain structures that will be generalized to
systems with broken symmetries with and without potentials. 
Even more symmetry breaking of non-compact groups will be considered. 

\subsection*{Simple Examples for Symmetry Breaking in General}
There are many examples known in physics which cause spontaneous symmetry 
breaking by 
potentials of the form (\ref{HPot}) or (\ref{MHPot}), e.g.~ferromagnetism, 
superconducting and superfluidity. We like to give some purely structural  
examples of symmetry
breaking in a more general sense. Hence, due to the missing of dynamics, 
no massless Goldstone excitations will appear. However these examples are
examples of symmetry breaking in general,   
mostly not familiar in connection
with symmetry breaking, but rather trivial. Not in all
cases a potential is needed. The same effect of breaking a global symmetry may  
equally be caused by initial conditions in a dynamical problem. In both
structures, spontaneous symmetry breaking caused by the ground state of a
potential and symmetry breaking by the initial conditions, the basic
ingredient is the distinction of a non-trivial vector in the
representation space of a group. 

In the following we like to give some really simple examples to make clear the
basic concepts of symmetry breaking within our familiar 
3-space $\Spa_3 \cong \R^3$. The symmetry group of
3-space is the 3-dimensional Euclidean group $E(3) \cong O(3) \times_s \Spa_3$ 
(here $\times_s$ denotes the semi-direct group product). The orbit of every
vector within $\Spa_3$ generated by the action of $E(3)$ is the whole 
vector space
$\Spa_3$ due to the affine normal subgroup $\Spa_3$. However, $\Spa_3$ is 
decomposed 
in two different strata with respect to the action of $O(3)$. The trivial
stratum is given by the origin of $\Spa_3$ and the non-trivial stratum includes
every non-trivial vector. The little groups of the non-trivial vectors in
$\Spa_3$
are isomorphic to O(2). Therefore every non-trivial
orbit is isomorphic to $O(3)/O(2)$ which itself is isomorphic to the 
2-sphere\footnote{in general $O(n+1)/O(n) \cong \$^n$} $\$^2$. 
With this simple mathematical background
one can imagine several examples of breaking O(3) down to O(2) 
whenever a non-trivial vector is given in 3-space. 

Take a ball, which is the prototype of a $O(3)$-symmetric
object. When acting with an $O(3)$-rotation on the ball nothing changes. 
However,
when acting with a non-trivial translation $t \in \Spa_3$ onto the ball, 
i.e.~the
ball is moving, this movement $t$ defines a
non-trivial vector in $\Spa_3$. The residual symmetry group of this system is 
the
little group of $t$, isomorphic to $O(2)$. The direction of the movement can be
characterized by a dot on the surface of the ball which would not change its
location when acting with $O(2)$ onto the ball. All possible directions are
given with all possible dots on the surface of the ball 
which is the 2-sphere $\$^2 \cong O(3)/O(2)$. Here 
the breaking of the symmetry is due to the initial conditions. 

The breaking of $O(3)$-symmetry of 3-space to $O(2)$ is even quite closer 
to us. 
Imagine the earth 
with its gravitational attraction. This defines a $O(3)$-symmetric system.
Now imagine you are going down to earth e.g.~with a
space ship. You have to set down on a single point on earth.  
However, now the $O(3)$-symmetry is `lost'. Bound to the earth's surface by the
gravitational attraction everybody defines a non-trivial vector from the center
to the surface
of the earth. We individually have no freedom of the whole $O(3)$-symmetry
of 3-space, 
but feel only the residual $O(2)$-symmetry of the tangent 
space onto the earth's surface (e.g.~when dancing on the floor), defined by 
the non-trivial vector from the center of the earth to ourself. However, when 
losing the global symmetry we gain the new, but local concepts of 
`up'\footnote{The absolute
`up' direction in say Australia and Europe may be quite different, however, no
misunderstandings would arise.} and `down'. 
Hence, symmetry breaking not only reduces the symmetry of the system but at
the same time creates new structures by decomposing the representation space.  
This decompositions can locally be interpreted as vector
space decompositions. In our example it is a decomposition of $\Spa_3$ into the
tangential plane and the orthogonal line (up \& down) 
due to the one distinguished (basis)
vector\footnote{In general a complete basis of a vector space defines a
complete decomposition of the vector space with respect to its field and
therewith an isomorphism, e.g.~$\Spa_3 \cong \R \oplus \R \oplus \R \cong 
\R^3$.}, 
$\Spa_3 \cong \R \oplus \Spa_2$, with $\Spa_2 \cong \R^2$. 
This decomposition is only given as local tangent space onto the global
structure. 
The global structure however 
is a manifold decomposition of the orbits within the strata, i.e.~$\Spa_3 \cong
D(1) \times O(3)/O(2) \uplus \{ 0 \}$. Therefore motion on the `Goldstone
manifold' O(3)/O(2), i.e.~on the earth's surface, takes no (potential) energy,
in analogue to the massless Goldstone excitations. On the other hand the $D(1)$
direction (up \& down) is the analogue to the massive Higgs 
excitation\footnote{Notice, that it is the gravitational
potential which is the prototype for all other potentials and thus also for the
Higgs potential. These analogies are therefore not astonishing but quite
natural.}. 
This quite comprehensible example  
provides thus all properties of spontaneous symmetry breaking.  

One remark to our local $O(2)$-symmetry: How can we see the relevance of this
residual symmetry for us individually? We are not $O(2)$-symmetric, whereas 
plants, bound to a single point on earth, are (more or less)
$O(2)$-symmetric, best example may be given by a fir. For a tree there is 
normally no
reason to prefer one direction. This is different for all kinds
of moving objects on earth (biological or artificial). The possibility to
change the location again breaks the $O(2)$-symmetry down to discrete group of
space reflections denoted by $\I_2$. Hence, we and all other moving objects are
(again more or less) symmetric under reflection of the space. Therefore we
can distinguish our local surroundings in what's ahead and what's behind us. 
It is harder to define what's right and what's left.  

Back to the example of symmetry breaking of every individual on earth. We are
forced to live on the earth's surface by the $1/r$ gravitational potential of
the earth on the one side and the resistance of the matter of the earth on the
other side. Hence, everybody (and every body on earth) 
defines one different $O(2)$-symmetric ground state 
in this $O(3)$-symmetric potential. This is like spontaneous symmetry breaking
in the Higgs mechanism. However, for the system earth there is also a breaking
of the $O(3)$-symmetry down to $O(2)$ given by dynamics, i.e.~by the
rotation of the earth itself, defining the north-south axis. 
Does there arise a problem when two different structures of symmetry breaking
occur? Locally this seems to be not any problem. However, when moving on
the earth's surface, e.g.~from north to south, there arise Coriolis forces 
due to the
rotation of the earth. These Coriolis forces vanish when travelling according 
to the
symmetry of the rotation axis, i.e.~according to the little group of the axial
vector. 

We have shown the most basic concepts connected with symmetry breaking in 
quite vivid examples. We can now take the road to more abstract systems
in which symmetry breaking occurs.  

\subsection*{Massive Objects}
In the example given above symmetry breaking was considered in 3-space 
with the action of
$O(3)$. This system is embedded in Minkowski space with the action of the
Lorentz group $O(1,3)$. In 1908 Hermann Minkowski began his famous talk in 
front of 
the `Gesellschaft der \"Arzte und Naturforscher' with the words:
`Von Stund an sollen Raum f\"ur sich und Zeit f\"ur sich v\"ollig zu Schatten
herabsinken und nur noch eine Art Union der beiden soll Selbst\"andigkeit
bewahren'\footnote{From this hour on space itself and time itself should vanish
into shadows but
only a kind of union of both will regard independence.} \cite{Mink}. 
Notice, when using the word `from now
on' Minkowski himself distinguished between time and space (he didn't say 
`from now and here on'). Why are for us individually
space and time distinguished? Again this could be a matter of a local vector
space decomposition coming along with symmetry breaking. Like everybody on
earth defines a non-trivial vector to the earth's surface, everybody, and every
massive object, defines a massive vector $p$ in Minkowski space 
($p_\mu = (m, 0, 0, 0)$ in a certain basis). The little group
of this vector is isomorphic to $O(3)$, compatible with the local vector space
decomposition $\M \cong \T \oplus \Spa_3$. 
Let us stress this again: Every massive object separates the action of the
Lorentz group into the action of the space rotations $O(3)$, the little group 
acting trivial, and the Lorentz boosts $O(1,3)/O(3)$. This provides a local
decomposition of Minkowski space into time and space. 
Unfortunately, unlike in the above
example, where we in principle can leave earth to feel the whole 
$O(3)$-freedom, we can't get out of this
decomposition to experience the whole Minkowski space, if there is any linear
space-time, and to feel the whole freedom of the Lorentz group. 

The embedding of this local or individual space-time  
decomposition $\T \oplus \Spa_3$ into Minkowski space $\M$ is given by 
special relativity. 
However, the global decomposition coming along with a massive vector is the 
manifold
decomposition of the interior part of the forward lightcone, the stratum of the
massive vector, into the Lorentz hyperboloids given with the
manifold $D(1) \times SO^+(1,3)/SO(3)$. 
Here again the non-trivial vector $p_\mu$ defines one orbit
within the time-like stratum breaking the $D(1)$ structure of the stratum 
and introducing a scale. The tangent space on 
every point of this manifold is the Minkowski space in a specific space-time
decomposition. This manifold can therefore be regarded as a prototype for a 
space-time
manifold \cite{Sal 97}, i.e.~locally Minkowskian, generated by a massive
object. Apart from a potential, all structures of symmetry breaking are 
present in this example. 

Our individual left-right-symmetry is thus the
result of a pattern of symmetry breaking down from the Lorentz group: 
\begin{itemize}
\item Being massive we define a local space-time decomposition in Minkowski
space, $\M \cong \T \oplus \Spa_3$, providing the concept of time itself and
`at the same time' the freedom of 3-space with its $O(3)$-symmetry. 
\item Bound to earth we define our local flat surroundings with 
$O(2)$-symmetry, the tangential plane, and the orthogonal direction, 
$\Spa_3 \cong \Spa_2
\oplus \Spa$. For us, being no birds, the vertical direction is quite different
to the tangential plane like time is different to space. 
\item The possibility to move defines the concept of `ahead' and `behind',
depending on the direction of our (potential) motion, $\Spa_2 \cong \Spa \oplus
\Spa \cong \R \oplus \R$, with residual reflection symmetry. 
\end{itemize}
Evolution built our
bodies (with some exceptions like the location of our heart) according
to this symmetry structure. How could it have done different? 
   
\subsection*{Wigner Classification in Terms of Symmetry Breaking} 
The example of a massive vector in Minkowski space is one part of the Wigner
classification of particles \cite{Wig PG} interpreted in terms of symmetry
breaking. Particles are, according to Wigner, positive unitary irreducible
representations of the Poincar\'e group. Therewith particles and their
properties are classified due to the different $O(1,3)$ strata\footnote{For
simplicity we don't pay any attention to the discrete structures in $O(1,3)$.
The structures connected with the reflections in $O(1,3) \cong \I_2 \times_s 
(\I_2 \times SO^+(1,3))$ and the discrete 
symmetries in quantum field theories (T, CP, CPT) are treated elsewere
\cite{Mah}.} 
in Minkowski space with momentum vector $p$ being 
\begin{enumerate}
\item time-like, 
\item light-like,
\item vanish, 
\item space-like.    
\end{enumerate}

The 1st class - our previous example of massive particles - is characterized by
real non-zero mass and discrete spin due to the spectrum of the compact little
group $SO(3)$ for massive vector 
bosons (spin $1$ with 3rd components $\pm 1, 0$) 
or its covering group $SU(2)$ for massive fermions 
(spin $1/2$ with 3rd components $\pm 1/2$). 
Compatible with the action of this
little group is the local space-time decomposition $\M \cong \T \oplus \Spa_3$
(or Sylvester decomposition).    

The 2nd class, the stratum of zero-mass but non-vanishing vectors, is
characterized by little groups isomorphic to $E(2)$. 
However, the vector space
decomposition into one light-like direction $\Li$ is not compatible with the 
action of this little group, since the direct complement of $\Li$ is not
invariant under the action of $E(2) \cong SO(2) \times_s \R^2$. 
I.e.~$E(2)$ acts not irreducible on this
decomposition, the action of $\R^2 \subset E(2)$ leaves
invariant the vector space $\Li$, but mixes this direction into the direct 
complement. Only two linearily independent light-like vectors define a vector
space decomposition \cite{SBH}, $\M \cong \Li_+ \oplus \Li_- \oplus \Spa_2$ 
(Witt decomposition), out of which a
space-time decomposition (Sylvester decomposition) 
can be constructed, $\Li_+ \oplus \Li_- \oplus \Spa_2
\cong \T \oplus \Spa_1 \oplus \Spa_2$. 
The little groups of this decomposition - 
in this case it is no fixgroup of a vector, but the stability group of the
decomposition - are isomorphic to $SO(2) \cong U(1)$.   
The spectrum of this little group is discrete as can be seen by the 
polarisation 
states of the photons or, if massless, the helicity states of the neutrinos.  

The 3rd class is the trivial one with little group being identical with the
whole group $O(1,3)$. There are no particles connected with this
class. 

The 4th class is characterized by imaginary mass. The little group of a
space-like vector is isomorphic to the non-compact group $SO(1,2)$ with
continuous spectrum. Therewith this class is characterized by continuous spin,
however, there are again no particles connected with this
class\footnote{Influenced by experiments on the spectrum of the beta decay of
tritium, which gave negative mass squared for the neutrinos, some people
speculated about neutrinos being tachyons \cite{Ntach}. 
If the Wigner classification due to
the Poincar\'e group is correct, tachyons would have continuous spin, which 
for neutrinos contradicts the experimental verified helicity strucure. Only
with the distinction of an additional vector, which breaks the little group 
$SO(1,2)$ again down to $SO(2)$, there would arise particles with discrete
spin.}. 

The manifold decompositions are
decompositions of the strata with the pattern $D(1) \times
SO^+(1,3)/H$ with $H$ denoting the little group, 
whereas the vector space decompositions, linear
but local, recover the whole Minkowski space. Again this is consistent with
our experience of space: We receive no direct knowledge from 3-space. All our
information comes from within our current backward lightcone. 3-space is only 
a linear extrapolation, given together with the assumption, that the 
information comes from
objects space-like to us in previous times\footnote{Maybe this is a rather
natural assumption, however, it is an assumption.}. 
I.e.~the linear Minkowski space is
only a linear reconstruction of space-time. Prior is the Lorentz orbit
structure or an even more complicated Einstein space-time. 

One remark to the classification of particles according to the Poincar\'e
group: Due to the affine space-time translations within the Poincar\'e group
particles are regarded to be asymptotic, i.e.~far of from (point like?)  
`interactions'. The
properties of the Poincar\'e group caracterized by the Casimir invariants of 
the Poincar\'e group, mass and spin, are thus properties of the asymptotic
particles. As can be seen in the standard model of elementary particles the
`particles' in the interaction are characterized by the Casimir invariants of 
only the Lorentz group or their covering group $SL(2,\C)$, 
i.e.~by chirality as expressed with left- and right-handed Weyl-spinors.  
These invariants are different to the Casimir invariants of the Poincar\'e 
group. 
Especially mass, connected with the translations, is no
invariant of the Lorentz group and thus no good
invariant of the interaction, but has to be renormalized\footnote{Keep in mind, 
that the renormalization of mass can be seen as a representation of 
$D(1) \cong \R^+$.}. 
The same seems to be
true for charge. Again it is the Higgs sector in the standard model which
connects the chirality structure of the Weyl spinors with the spin structure of
the Dirac spinors and at the same time 
generates mass and charge. Therefore the Higgs mechanism may be seen as the
link between the structures in the interaction and the asymptotics parametrized
with particles. 

\subsection*{Some Further Aspects of the Higgs Sector}
The Higgs potential is used to motivate a
non-trivial vector in the representation space $V \cong \C^2$ of an  
$U(2)$-symmetry. It is the Higgs potential in which the introduction of a 
scale is hidden. 
Apart from the already mentioned symmetry structure the non-trivial Higgs 
vector
defines locally, i.e.~on the Goldstone manifold $U(2)/U(1)$, a vector space
decomposition with one `direction' (the `charge direction') 
having the residual
$U(1)$-symmetry. This is parametrized with the Higgs field as follows:

The Higgs field $\phi(x) \in \C^2$ in its `symmetry breaking phase' is 
linearily expanded from the non-trivial Higgs vector, e.g.~$\phi_0 = {\tiny 
\left(\begin{array}{c} 0 \\ v \end{array}\right)}$, in cartesian coordinates, 
regarded as infinitesimal excitations,    
\beq
	\phi(x) \ = \ \phi_0 + \left(\begin{array}{c} 0 \\ \eta(x) 
		\end{array}\right) + \left(\begin{array}{c} \phi_1(x) + 
		i \phi_2(x) \\  i \phi_3(x) \end{array}\right) \ \in \C^2 .
\end{equation} 
The $\phi_i(x)$ are the Goldstone fields (excitations in directions of
the ground state manifold) and the $\eta(x)$, the excitation in
direction of the Higgs vector, becomes the massive Higgs particle. This is the
linear expansion of $\C^2$ out of one non-trivial vector. It is 
according to the
`tangent space decomposition' $\C^2 \cong \R \oplus \R^3$. This 
tangent space decomposition is regarded to be done at the
manifold decomposition of the stratum induced by the non-trivial Higgs vector: 
$\C^2 \cong D(1) \times U(2)/U(1)
\uplus \{0\}$. The manifold decomposition is used for the Higgs field expansion
in polar coordinates, adapted to the symmetry structure of the
problem\footnote{For simplicity we here used a representation of $U(2)$ rather 
than of $U(2)/U(1)$. Hence there is one `superfluous parameter'. 
E.g.~for $\phi_0$ given in
a certain basis by $\tiny \left( \begin{array}{c} 0 \\ v \end{array} \right)$
the action of $\1_2 + \tau_3$ is trivial, giving the residual charge-$U(1)$
representation. Therewith in this basis the combination
$\tilde{\phi}_0(x) + \tilde{\phi}_3(x)$ is no parameter of the Goldstone
manifold.}: 
\beq
	\phi(x) \ = \ e^{\tilde{\eta}(x)} 
		e^{i (\tilde{\phi}_i(x) \tau_i + \tilde{\phi}_0(x) \1_2)} 
		 \ \phi_0 . 
\end{equation}
It is the dilatational excitation, $D(1) \notin U(2)$ but $D(1) \in
GL(2,\C)_{\sR}$, 
which is connected with the massive Higgs particle\footnote{The groups 
$GL(2,\C)$ and $SL(2,\C)$ are
regarded in the following as real Lie groups.}. 

Via the Yukawa couplings in the Weinberg-Salam model 
the properties of the Higgs sector are carried to the electron-neutrino 
vector space. I.e.~one
aspect of the Higgs sector is to define, within the 
$U(2)$-representation space of the electron-neutrino\footnote{$U(2)$
and $U(1) \times SU(2)$ have the same Lie algebra $u(1) \oplus su(2)$ 
because of the isomorphism $U(2) \cong (U(1)
\times SU(2))/\I_2$. Due to the common discrete factor the representations of
$U(2)$ are representations of $U(1) \times SU(2)$ with a specific
correlation. The particle spectrum of the standard model shows exactly this
correlation \cite{OR,HS}. Therefore we regard $U(2)$ to be the symmetry group 
of the electro-weak interaction.},   
the `local' difference between neutrino and electron, with residual non-trivial 
charge symmetry $U(1)$ for the electron. 
To be more precise: Symmetry means `there is no difference' when acting with
acertain operation and so there is no difference between electron and 
neutrino in what
we call `interaction'. However, asymptotically we distinguish between electron 
and neutrino, both having different mass and charge. Therefore we have to
distinguish between the electron-neutrino-field in the interaction and the
electron and the neutrino in the asymptotics. This difference is 
defined by the non-trivial Higgs
vector, leading to the concept of mass and charge for the particles.    

Now at the latest the question arises, whether
this needs to be caused by a potential or whether there may be another 
structure like in the Wigner classification. Let us face some more questions in
this context: The Higgs mechanism introduces mass terms in the standard
model Lagrange density in a gauge invariant way.
On the other hand mass is defined as Casimir invariant
of the Poincar\'e group. The later concept is used in the Wigner classification
of particles, which we have seen can equally be interpreted in terms of
symmetry breaking. Thus we have at least two concepts of mass: the `Higgs mass'
and the `Poincar\'e or Wigner mass'. 
These concepts of mass are uncorrelated 
like it was the case for gravitational mass and inertial mass before general 
relativity. Could one believe that 
there are two different structures in special relativistic
quantum field theory (omitting even general relativity with quite another
aspect of mass) leading to the same concept of mass? Isn't it astonishing that
both structures can be interpreted in terms of symmetry breaking? Shouldn't it
be only one mechanism which generates the internal and external aspects of
mass, i.e.~only one symmetry breaking vector? Could it be that the $D(1)$
structures for massive particles in the Wigner case and in the Higgs case are
quite the same, i.e.~the massive vector in Minkowski space and the non-trivial
vector for the Higgs mechanism are of the same origin? 
We will in the following section present a first attempt for a model
based on the Higgs mechanism and on the space-time model of Saller \cite{Sal
97} which will generate at least the symmetry structure of the Weinberg-Salam
model and of a massive particle with only one symmetry breaking vector. The
equality of `Higgs-mass' and `Wigner-mass' are thus evident per construction.

\subsection*{A Unified Model}
Let us sum the structures of mass connected with symmetry breaking: 
On the one
hand we have the external structure of massive objects decomposing the
time-like stratum $D(1) \times SO^+(1,3)/SO(3)$ with broken (or fixed) $D(1)$
scale invariance and residual symmetry $SO(3)$. On the other hand we have the
internal structure decomposing $\C^2$ according to $D(1) \times U(2)/U(1)
\uplus \{0\}$ again with fixed $D(1)$ scale but residual 
symmetry $U(1)$. How do these structures fit together by identifying the
dilatation structures? Notice the observation in \cite{Sal 97}: 
\beq
	SO^+(1,3)/SO(3) \ \cong \ SL(2,\C)/SU(2)  \ \cong UL(2,\C)/U(2) ,  
\end{equation}
with $UL(2,\C) := \{g \in GL(2,\C) | \abs{\mbox{det} g} = 1 \} 
\cong GL(2,\C)/D(1)$ the group of linear  
operations with determinant of modulus one\footnote{This group will be named
according to \cite{Sal 97} 
unimodular linear group. In general 
$UL(V) := \{g \in GL(V) | \abs{\mbox{det} g} = 1 \}$. 
For every finite dimensional vector space $V$ we have the direct group
product separation: $GL(V) \cong D(1) \times UL(V)$.}. Hence we should  
analyze the action of the whole $GL(2,\C)$.  

Imagine the action of $GL(2,\C)$ on any non-trivial vector $v$ in 
$V \cong \C^2$:
There is no natural basis in $V$. Thus this vector defines one (normal) 
basis vector and
all other vectors in $V$ can be `measured' with respect to $v$ if there is in
addition a bi- or sesquilinear form. Hence together with a positive definite
sesquilinear form (which is equivalent to a positive definite and therewith 
$U(2)$-invariant conjugation on $\C^2$) it
defines a `length' or `energy' scale in $V$. In $GL(2,\C)$ the dilatations 
are separated naturally according to the direct group product 
$GL(2,\C) \cong D(1) \times UL(2,\C)$.   
However, the action of $GL(2,\C)$ on a non-trivial
vector $v \in \C^2$ together with a positive definite conjugation 
decomposes this group according to 
\beq
	GL(2,\C) \bullet v \ \cong \ 
		\left( D(1) \times UL(2,\C)/U(2)
		\times U(2)/U(1) \times U(1) 
		\right) \bullet v. \label{RZMZerl}
\end{equation}

We
have to emphasize that this is a manifold decomposition, no direct group product
decomposition. The later part of this decomposition 
has the symmetry structure of the Higgs sector in the Weinberg-Salam model: 
$U(1)$ is the little group of the non-trivial vector $v$ acting trivial. The
action of 
$U(2)/U(1)$, being the Goldstone manifold $[v]_{U(2)}$, generates vectors in
$\C^2$ which have the same `length'\footnote{We used a
positive devinite conjugation to define the group $U(2)$ within $GL(2,\C)$ and
therewith a positive definite sesquilinear form providing the concept of a
positive definite (length or energy) scale. We could have equally used the
indefinite conjugation on $\C^2$ with $U(1,1)$ invariance group. This would
also introduce a sesqilinear form for measuring vectors in $\C^2$ with respect
to $v$, 
but with indefinite results. I.e.~it would introduce a scale, but not 
interpretable in general as energy scale.
The decomposition would be according to 
$GL(2,\C) \cong D(1) \times UL(2,\C)/U(1,1) \times U(1,1)/U(1) \times U(1)$.}  
and thus introducing the same length or energy scale. 
The fixgroups of the elements of this coset space are
isomorphic but not identic. I.e.~$U(1)$ is local on the Goldstone manifold. 
The action of the coset 
\[
	UL(2,\C)/U(2) \cong SL(2,\C)/SU(2) \cong SO^+(1,3)/SO(3) 
\] 
on the non-trivial
vectors in $[v]_{U(2)}$ is trivial. Regarded as manifold it is
isomorphic to the Lorentz orbit of a massive particle. I.e.~a Lorentz boost
doesn't change the (energy) scale, but
defines equivalent space-time decompositions. 
Moreover, 
$D(1) \times UL(2,\C)/U(2)$ is isomorphic to the stratum of massive particles,
the interior part of the (forward) lightcone, with tangent space being the
Minkowski space \cite{Sal 97} in its local space-time decompositions. Hence, we
have the non-linear space-time structure for a massive object on the one 
side \cite{Sal 97}, 
\[
	D(1) \times UL(2,\C)/U(2) \cong GL(2,\C)/U(2),
\] 
and the symmetry structure of the Higgs sector
in the standard model on the other side \cite{Wei}, 
\[
	U(2)/U(1) \times U(1) \cong U(2),
\] 
giving together the action of $GL(2,\C)$: 
\beq
\begin{array}{ccccc}
	GL(2,\C)  \ \cong \ 
		\left( \right. & \underbrace{D(1) \times UL(2,\C)/U(2)} &
		\times  &\underbrace{U(2)/U(1) \times U(1)} &
		\left. \right)   \\
	& \mbox{space-time} & & \mbox{Higgs sector} & 
\end{array} \nonumber
\end{equation}
Notice that the action of $U(2)$
again is local on the space-time manifold $GL(2,\C)/U(2)$, i.e. the action of
$U(2)$ is space-time dependent. 

The decomposition of $GL(2,\C)$ according to (\ref{RZMZerl}) is no group 
product decomposition. I.e.~the group $GL(2,\C)$ itself acts not irreducibly on
this decomposition. Only supgroups of $GL(2,\C)$ have the possibility to act
irreducible on parts of this decomposition. Hence the action of $GL(2,\C)$ on
the manifold decomposition (\ref{RZMZerl}) will be quite non-trivial, resulting
in a new isomorphic, but different decomposition. 

What makes this simple model of the action of $GL(2,\C)$ on a non-trivial
vector worth being considered? First of all, one has only one symmetry breaking
mechanism generating the internal and external symmetry structure for massive
particles and introducing only one mass or energy scale. Higgs-mass and
Wigner-mass are identical a priori. 
Moreover, the generation of external and
internal symmetry structures (besides the symmetry structure for quarks) `at
the same time' may lead to the occurence of space-time and matter
(particles, objects) on equal level: Since Newton we organize objects and 
their motion in our space-time, regarded space-time being a priori. This is 
like distributing objects in a given box. 
General relativity relates the geometry of
this box to the distribution of its contents, the box itself (the Einstein
manifold) stays a priori. 
This results in the difference of geometry on the one side and matter on the 
other side in the Einstein equation, a duality which Einstein himself always 
tried to resolve. 
But how do we experience space-time? According to Mach's principle one can only
define energy or mass, momentum, and angular momentum with respect to other 
objects. The same is true for length or volume. Even more, we can test 
space-time itself only with the help of its `contents'. 
According to the philosophy of Leibniz and
Mach space-time can only be regarded as relation between objects,
i.e.~space-time without contents doesn't exist at all. But what
was first, space-time or matter? This is like the problem of the hen and the
eg. The solution of this paradoxy is, that none
`was first', but both structures are generated `at the same time',
i.e.~space-time only comes along with matter and matter only comes along with
space-time. The above model generates external and internal
symmetries out of one root 
and therewith has the potential to generate space-time and matter 
`at the same time'. This is done with the help of
the concept of symmetry breaking given in a more general framework.

\subsection*{Outlook}
In the context of the structures suggested above there immediately arise some
questions. Let us mention some of these questions and hints to adress them 
instead of a conclusion: 

Where does the non-trivial vector $v$ come from when not spontaneousely chosen
as ground state of a potential? In quantized theories it is not the Higgs
vector which is non-trivial, but the vacuum expection value of the Higgs field.
Hence, when there seems to be a symmetry breaking ground state it could be the
`vacuum' of a Fock space in a quantized system out of which the whole Fock
space will be generated. The calculation of the spectrum of a quantized theory
of $\C^2$ with non-invariant Fock ground state is possible 
together with the canonical 
quantization of a vector space $V \cong \C^n$ in the basis-free
formulation according to 
H.~Saller \cite{Sal 87}. The quantization of $\C$ gives just the quantum
mechanical harmonic oscillator with its spectrum. 
The quantization of $\C^2$ with non-trivial 
Fock ground state will generate a spectrum of `particles' which 
at least will show the symmetry structure of (\ref{RZMZerl}).

On the other hand we up to now only dealt with one symmetry breaking vector.
Which structures will arise when there are two different vectors with
isomorphic, but different separations according to (\ref{RZMZerl})? Is there
the possibility of an even more complicated Einstein manifold instead of the
foreward time-like cone structure, generated out of more than one non-trivial
vector? The example
of the earth hints that there is the possibility of `Coriolis forces' when
breaking the symmetry in different manners. Could
gravity be the result of generalized
Coriolis forces caused by two or more different symmetry breaking structures?  
dynamical,

Two different vectors in $\C^2$ are connected by the action of a distinct
element $g \in GL(2,\C)$. 
But how does the action of $g \in GL(2,\C)$ change the manifold decomposition
(\ref{RZMZerl})? And how would we interpret this isomorphic, but different
decomposition and therewith the action of $GL(2,\C)$ itself? 

If we generate the
concept of mass and energy out of deeper structures via symmetry breaking, what
is the concept of dynamics and therewith how do thus equation of motions
(strongly correlated to the concept of time) or
Lagrange densities (strongly correlated to the concept of energy)
evolve?  

We only considered purely structural concepts. Which aspects will
arise when making this system dynamical? Does there appear Goldstone modes
together with the two cosets $U(2)/U(1)$ and $UL(2,\C)/U(2) \cong
SO^+(1,3)/SO(3)$ and what will be their role especially in the case of the
non-compact coset? Is there a connection to the geometry on this coset
space? 

\section*{Acknowledgment}
I would like to thank H.\,Saller for the profound discussions on the 
structures of space-time and matter and B.\,Fauser for helpfull advices 
as well as C.S.


\end{document}